# Title: FOUNDATIONS FOR AI-ASSISTED FORMATIVE ASSESSMENT FEEDBACK FOR SHORT-ANSWER TASKS IN LARGE-ENROLLMENT CLASSES

**Authors:** Susan E Lloyd, Matthew D Beckman, Dennis K Pearl, Rebecca J Passoneau, Zhaohui Li, Zekun Wang
**Affiliation of Author(s):** The Pennsylvania State University
**Single Contact Email Address:** mdb268@psu.edu

## Abstract

Research suggests "write-to-learn" tasks improve learning outcomes, yet constructed-response methods of formative assessment become unwieldy with large class sizes. This study evaluates natural language processing algorithms to assist this aim. Six short-answer tasks completed by 1,935 students were scored by several human raters, using a detailed rubric, and an algorithm. Results indicate substantial inter-rater agreement using quadratic weighted kappa for rater pairs (each QWK > 0.74) and group consensus (Fleiss' Kappa = 0.68). Additionally, intra-rater agreement was estimated for one rater who had scored 178 responses seven years prior (QWK = 0.89). With compelling rater agreement, the study then pilots cluster analysis of response text toward enabling instructors to ascribe meaning to clusters as a means for scalable formative assessment.

## Introduction

Effective formative assessment is indispensable for students and instructors to monitor learning (GAISE, 2016; Pearl, 2012). Furthermore, it is critical for a citizen statistician to be able to communicate statistical ideas effectively, both as a consumer and a producer of statistical information (Gould, 2010). One avenue through which students develop these effective communication skills is through written tasks. In fact, research has linked "write-to-learn" tasks to improved learning outcomes in science and mathematics, yet constructed-response methods of formative assessment such as minute papers and comprehension questions become unwieldy for instructors with large class sizes (e.g., hundreds, thousands) (Woodard, 2021). A human-machine collaboration may provide the means necessary to improve the feasibility of formative assessment at scale as well as the quality of feedback provided to large enrollment students (Basu, 2013). In the current literature, AI-assisted formative assessment feedback has primarily only been presented for essays or long-answer tasks, and in disciplines other than statistics (see e.g., Attali, et al., 2008; Page, 1994). This study serves as the groundwork for leveraging natural language processing (NLP) algorithms to assist formative assessment using short-answer tasks in large enrollment courses.

## Literature Review

Effective assessment feedback should be timely (GAISE, 2016; Garfield, 2008). Popular solutions for large enrollment classes often rely upon selected-response tasks (e.g., multiple choice) as a vehicle for formative assessment. For example, the GAISE (2016) guidelines recommend clickers and similar student response systems, coupled with engagement strategies to encourage careful reflection before and after responding, as a means for scalable formative assessment. Even still, selected-response formats tend toward lower levels of Bloom's Taxonomy such as recall and recognition tasks (Bloom, 1956; Garfield, 2008; Basu, 2013). The format also invites guessing, which impairs the instructor's ability to differentiate between the demonstration of the desired learning outcome as opposed to a lucky guess, leading question, or ineffective distractors (Jordan, 2009). By comparison, short-answer response tasks allow students to articulate their reasoning and have greater potential to invoke higher levels of thinking on Bloom's Taxonomy (Theobold, 2021).

When students reason and communicate through writing, it serves as a vehicle for sharpening understanding (Graham, et al., 2020). Continual practice with communicating statistical information, ideas, and thinking in this manner is thought to improve statistical literacy and learning outcomes as well as promote retention (Basu, 2013). Such tasks enable students to explain concepts, justify conclusions, apply knowledge to new scenarios, and form disciplinary connections in their own words (Bloom, 1956; Garfield, 2008; Graham, 2020). Students with varying degrees of correctness and understanding warrant different types of feedback (Basu, 2013; Jordan, 2009). Short-answer response tasks also allow instructors to more easily identify student misconceptions and address student misunderstandings that may otherwise have gone undetected (Basu, 2013). In this way, instructors can more closely monitor students' learning and understanding, resulting in effective formative assessment (GAISE, 2016; Pearl, 2012).

A human-machine collaboration is a promising mechanism to assist rapid, individualized feedback at scale (Basu, 2013). Natural language processing (NLP) methods can achieve reliable classification (e.g., incorrect / partial / correct) of short-answer responses, which could be followed by automatic clustering of similar student responses for formative assessment. Reliable classification means the algorithm assigns appropriate scores to the responses, aligning with the pre-established scoring reliability metrics. Successful clustering would group student responses into clusters that are as homogenous within, and as heterogeneous between, as possible. The objective would be to iteratively refine the clustering so an instructor can attach meaning to clusters of responses (Basu, 2013). By exploiting the efficiency of technology for short-answer tasks, students in large enrollment classes can access a type of timely, personalized feedback believed to enhance the learning experience in smaller classes (Basu, 2013; Wright, 2019).

Scoring reliability is the broad term for assessing the consistency with which raters score, or label, a given response. Inter-rater reliability refers to comparing the reliability of scores among one or more trained human raters, while intra-rater reliability refers to comparing the reliability of scores from one human rater at two different points in time (Gwet, 2008). With the emergence of automated rating systems, an algorithm can serve as one of the trained raters being considered in a scoring reliability analysis (Basu, 2013). An algorithm's reliability can be similarly scrutinized by comparing the reliability of a classification algorithm to that of human raters. Since human raters are fallible and prone to inconsistencies and biases, there is the need to establish a more reasonable standard of comparison aligned to the reliability expected of competent human raters when judging the performance of an algorithm (Woodard, 2021; Page, 1994).

Toward the goal of improving the balance between the instructor burden and student benefit associated with formative assessment, this research study aims to address the following questions: (RQ1) What level of agreement is achieved among trained human raters labeling (i.e., scoring) short-answer tasks? (RQ2) What level of agreement is achieved between human raters and an NLP algorithm? (RQ3) What sort of NLP representation leads to good clustering performance, and how does that interact with the classification algorithm?

**Methods**

This study utilized de-identified extant data from a previous study, which solicited responses to a group of short-answer tasks from post-secondary students enrolled in introductory statistics courses (Beckman, 2015). The data consist of responses to 6 short-answer tasks provided by 1,935 students representing a total of 29 class sections for 16 unique courses at 15 distinct institutions that are mostly, but not exclusively, located in the USA.

(RQ1) The 1,935 students from the 2015 study, and their associated responses to each task, were divided among four persons with sufficient intersection to evaluate rater agreement. The four persons possess varied levels of experience with statistics education that would be common within an instructional team. Rater A was an experienced statistics instructor and the author of the tasks' prompts and associated scoring rubrics. Rater B was an experienced statistics instructor. Rater C was a statistics graduate student with some experience as a teaching assistant in statistics and had previously taught an undergraduate mathematics course. Rater D was a statistics graduate student teaching assistant. The study sought to evaluate all student responses available, with quality responses from at least 50 students for the analysis of agreement between each possible combination of raters for RQ1.

Using a prior analysis to estimate the approximate proportion of earnest response attempts in the data, each desired rater comparison was allocated 63 randomly selected students to target approximately 50 quality responses. Therefore, three raters (i.e., Rater A, Rater B, Rater C) were assigned to review responses by 750 students such that each pair of raters would share an intersection of 63 randomly selected students in addition to a distinct set of 63 randomly selected students shared by all three raters. After the initial allocation exercise, but before the scoring process, a fourth evaluator (i.e., Rater D) joined the study team and was assigned the 252 students previously assigned for multiple raters (63 x 3 pairwise + 63 three-way).

The only constraint on the allocation of students to each rater was imposed to preserve a unique opportunity to examine intra-rater agreement for Rater A. Using the same rubric in service of an entirely different research objective, Rater A had scored a random sample of 178 responses in 2015 (see Beckman, 2015). The sample allocation to each rater in the present study simply verified that at least 50 of the students scored by Rater A in 2015 would again be evaluated by Rater A in the current study. Rater A had not revisited the scoring for those tasks during the 7 years elapsed.

Each evaluator used a detailed rubric to score the assigned student responses (see Beckman, 2015). Student responses were either given a score of 0: incorrect, 1: partial, or 2: correct, and examples of student responses for each classification were provided in the rubric. After all responses had been scored, confusion matrices were tabulated to determine the percentage agreement as well as the amount of one-level and two-level discrepancies. Scoring reliability among raters was estimated using quadratic weighted kappa (QWK) for pairwise agreement and Fleiss' kappa to measure consensus among three or more raters. Viera & Garrett (2005) describe a heuristic interpretation of rater agreement represented by various kappa values: kappa < 0 is worse than chance; 0 < kappa < 0.2 is slight agreement; 0.2 < kappa < 0.4 is fair agreement; 0.4 < kappa < 0.6 is moderate; 0.6 < kappa < 0.8 is substantial agreement; 0.8 < kappa < 1 indicates almost perfect rater agreement.

(RQ2) The scoring reliability measures for the four trained human raters served as a baseline with which to evaluate the algorithm performance and validate the reliability of automated scoring. For machine learning, the 7,258 unique task-responses were randomly split four ways: 90% were split into the typical division of training (72%), development (9%) and test (9%), with an additional 10% held in reserve for more rigorous testing. The 653 task-responses in the test set were selected to include responses with the highest agreement among human raters (e.g., 458 had unanimous agreement among 3 or 4 raters); the remaining task-responses were randomly assigned to the training, development, and reserve sets. Two NLP algorithms were compared for accuracy using a subset of student responses. The first being a logistic regression combined with a Long Short-Term Memory (LSTM) for learning vector representations, and the second being the Semantic Feature-Wise Transformation Relation Network (SFRN) (Li et al., 2021).

(RQ3) The goal of the clustering is to determine if a set of student responses that have the same correctness can be grouped into semantically similar clusters. The two NLP classification algorithms each learn a distinct vector representation on training data that supports better classification. Neither of these learned representations are optimal for clustering, which is a process to discover relationships in data, rather than to learn an *a priori* classification task. Therefore, we compare the clustering of the two types of learned vector representations with a third approach that applies a pre-trained phrase-embedding method to produce much lower dimension vectors. We compare all three using different clustering methods to develop insight into the best combination for semantic coherence of output clusters.

**Results**

(RQ1) When considering the inter-rater agreement among the three trained human raters (Raters A, C, & D), the pairwise quadratic weighted kappas (QWK) were between 0.79 and 0.83. The Fleiss' Kappa value for the three way comparison was 0.70 (see Table 1). These measures indicate substantial inter-rater agreement among the three human raters (Viera & Garrett, 2005). At the time of this writing, only data for tasks 2a and 2b could be evaluated for Rater B, but the results are similarly strong. The pairwise QWK between Rater B and other raters were between 0.71 and 0.74. The Fleiss' Kappa value for the four way comparison on tasks 2a and 2b was 0.62. When considering the intra-rater agreement for one evaluator, on a subset of the 178 responses scored from the study seven years prior, the pairwise QWK was 0.88. This measure indicates almost perfect intra-rater agreement following seven years elapsed (Viera & Garrett, 2005).

| Rater Comparison | Measure of Reliability |
| --- | --- |
| Rater A & Rater C | QWK = 0.8342 |
| Rater A & Rater D | QWK = 0.7966 |
| Rater C & Rater D | QWK = 0.7916 |
| Rater A (2015) & Rater A | QWK = 0.8802 |
| Rater A & Rater C & Rater D | Fleiss' Kappa = 0.698 |
| Rater A & SFRN | QWK = 0.7871 |
| Rater C & SFRN | QWK = 0.8151 |
| Rater D & SFRN | QWK = 0.7403 |
| Rater A & Rater C & Rater D & SFRN | Fleiss' Kappa = 0.678 |

Table 1: Reliability comparisons among human raters (A, C, D) and an NLP algorithm (SFRN).

(RQ2) Similar calculations were performed once the NLP algorithm was introduced as an additional rater. The SFRN algorithm achieved much higher classification accuracy than LSTM (83% vs. 72%). When considering SFRN and human raters, the QWK values for pairwise comparisons were between 0.74 and 0.82. The Fleiss' Kappa value for the four way comparison, between the algorithm and all three human raters, was 0.68 (see Table 1). Therefore, there was substantial inter-rater agreement among the raters, including the algorithm (Viera & Garrett, 2005). Other classifiers were tested but had much lower agreement.

(RQ3) SFRN learns a high-dimension (D=512) vector representation on training data, which as noted above produces high agreement with humans on a test set. Multiple experiments with K-means and K-medoids clustering of the test data showed that SFRN led to more consistent clusters when the representation is retrained (0.62), in comparison to other classifiers. Each class (correct, partially correct, incorrect) for each question is clustered separately. Consistency is measured as the ratio of all pairs of responses in a given class per question that are clustered the same way on two runs (in the same cluster, or not in the same cluster). However, the highest consistency (0.88; D=50) was achieved by generating a new representation for each response using WTMF (Guo & Diab, 2011), a matrix factorization method that produces static representations.

**Discussion**
**(1)Take-home message about the study**
In addition to laying the groundwork for NLP-assisted formative assessment feedback for short-answer tasks in large enrollment courses, this work presents a careful study of inter-rater agreement including varied experience typical for an instructional team of a large course, intra-rater agreement after seven years elapsed, and comparison between algorithm performance and domain experts using a detailed rubric. Given the high reliability of the algorithm, it's important to investigate how an environment could be created for teaching assistants and the algorithm to collaborate to achieve both high reliability on the scores and high quality feedback for students. The substantial scoring reliability and feasibility of clustering performance shown in this study suggest that a human-machine collaboration offers a promising opportunity for continued research toward a large class formative assessment using short-answer tasks that approaches small class quality and instructor burden.

**(2) Limitations of the study**
The study includes incomplete data for Rater B. The analysis does have data from Rater B with respect to two of the 6 tasks, but comparisons including Rater B are limited without the full data on the remaining tasks. In the 2015 study, students came from many classes of varying sizes, and not a single large class as desired. There is reason to believe this limitation would introduce noise into the data, likely resulting in conservative estimates of reliability and feasibility.
The key limitation of the NLP methods is how to manage the tradeoff between algorithms that achieve high reliability on classification of correctness based on neural network methods that learn high dimension vector representations, and the opposing requirement for low dimension representations to yield denser clusters with greater differentiation between clusters. Thus, we will pursue multiple avenues, such as dimension reduction prior to clustering, or a separate post-processing step that adopts an independent low dimension representation.

**(3) Implications for teaching and research**
There is intrinsic value in a rigorous evaluation of rater agreement for instructors of all class sizes. Investigating discrepancies and inconsistencies in scoring could lead to new insights regarding the nature of rater biases which could greatly contribute to the emerging area of data science ethics (Çetinkaya-Rundel 2021).
Although this study focuses on large enrollment classes in particular, success in these efforts creates an opportunity to study formative assessment interventions and mechanisms associated with desired learning outcomes that have implications for smaller and intermediate class sizes as well (Basu, 2013). For example, instructors of all class sizes would benefit by being able to focus their efforts on tasks other than grading, such as designing projects or studying how students respond to different types of feedback (Jordan, 2009). The use of an automated rater would also allow for the study of feedback effect and revision effect to determine whether students' learning experience is enhanced when given the opportunity to revise their responses (Attali & Powers, 2008).

**Acknowledgments**
The authors thank Joseph Miller for his contributions to scoring and conversations about the project.

# References

Attali, Y., & Powers, D. (2008). Effect of Immediate Feedback and Revision on Psychometric Properties of Open-Ended Gre® Subject Test Items. *ETS Research Report Series*, *2008*(1), i–23.

Attali, Y., Powers, D., Freedman, M., Harrison, M., & Obetz, S. (2008). Automated Scoring of Short-Answer Open-Ended Gre® Subject Test Items. *ETS Research Report Series*, *2008*(1), i–22.

Basu, S., Jacobs, C., & Vanderwende, L. (2013). Powergrading: a Clustering Approach to Amplify Human Effort for Short Answer Grading. *Transactions of the Association for Computational Linguistics*, *1*, 391–402. https://doi.org/10.1162/tacl_a_00236

Beckman, M. (2015). *Assessment Of Cognitive Transfer Outcomes For Students Of Introductory Statistics*. http://conservancy.umn.edu/handle/11299/175709

Bloom, B. S. (1956). *Taxonomy of educational objectives: The classification of educational goals*. Longman Group.

Çetinkaya-Rundel, M., & Ellison, V. (2021). A Fresh Look at Introductory Data Science. *Journal of Statistics and Data Science Education*, *29*(sup1), S16–S26.

GAISE College Report ASA Revision Committee (2016). Guidelines for Assessment and Instruction in Statistics Education College Report 2016. URL: http://www.amstat.org/education/gaise.

Garfield, J. B., Ben-Zvi, D., Chance, B., Medina, E., Roseth, C., & Zieffler, A. (2008). Assessment in Statistics Education. In J. B. Garfield, D. Ben-Zvi, B. Chance, E. Medina, C. Roseth, & A. Zieffler (Eds.), *Developing Students' Statistical Reasoning: Connecting Research and Teaching Practice* (pp. 65–89). Springer Netherlands. https://doi.org/10.1007/978-1-4020-8383-9_4

Graham, S., Kiuhara, S. A., & MacKay, M. (2020). The Effects of Writing on Learning in Science, Social Studies, and Mathematics: A Meta-Analysis. Review of Educational Research, 90(2), 179–226. https://doi.org/10.3102/0034654320914744

Gould, R. (2010). Statistics and the Modern Student. *International Statistical Review / Revue Internationale de Statistique*, *78*(2), 297–315. https://www.jstor.org/stable/27919839

Guo, W., Diab, M. (2012) Modeling Sentences in the Latent Space. In *Proceedings of the 50th Annual Meeting of the Association for Computational Linguistics*, pages 864–872. Association for Computational Linguistics.

Gwet, K. L. (2008). Computing inter-rater reliability and its variance in the presence of high agreement. *British Journal of Mathematical and Statistical Psychology*, *61*(1), 29–48.

Jordan, S., & Mitchell, T. (2009). e-Assessment for learning? The potential of short-answer free-text questions with tailored feedback: Short-answer free-text questions with feedback. *British Journal of Educational Technology*, *40*(2), 371–385. https://doi.org/10.1111/j.1467-8535.2008.00928.x

Page, E. B. (1994). Computer Grading of Student Prose, Using Modern Concepts and Software. *The Journal of Experimental Education*, *62*(2), 127–142.

Pearl, D. K., Garfield, J. B., delMas, R., Groth, R. E., Kaplan, J. J., McGowan, H., & Lee, H. S. (2012). Connecting Research to Practice in a Culture of Assessment for Introductory College-level Statistics. URL: http:// www.causeweb.org/research/guidelines/ResearchReport_Dec_2012.pdf.

Theobold, A. S. (2021). Oral Exams: A More Meaningful Assessment of Students' Understanding. *Journal of Statistics and Data Science Education*, *29*(2), 156–159.

Li, Z., Tomar, Y., & Passonneau, R. J., 2021. A Semantic Feature-Wise Transformation Relation Network for Automatic Short Answer Grading. In *Proceedings of the 2021 Conference on Empirical Methods in Natural Language Processing*, pp. 6030–6040. Association for Computational Linguistics. https://aclanthology.org/2021.emnlp-main.487

Viera, A. J., & Garrett, J. M. (2005). Understanding interobserver agreement: the kappa statistic. *Family Medicine*, *37*(5), 360–363.

Woodard, V., Lee, H., & Woodard, R. (2020). Writing Assignments to Assess Statistical Thinking. *Journal of Statistics Education*, *28*(1), 32–44. https://doi.org/10.1080/10691898.2019.1696257

Wright, M. C., Bergom, I., & Bartholomew, T. (2019). Decreased class size, increased active learning? Intended and enacted teaching strategies in smaller classes. *Active Learning in Higher Education*, *20*(1), pp. 51–62. https://doi.org/10.1177/1469787417735607